\documentclass[12pt]{article}
\usepackage{amsmath,amssymb,enumerate,amsfonts,mathrsfs,theorem}
\usepackage{graphicx}
\usepackage{subfigure}
\usepackage{color}
\usepackage{relsize}
\usepackage[retainorgcmds]{IEEEtrantools}
\usepackage{color}
\usepackage[framemethod=TikZ]{mdframed}
\usepackage{lipsum}
\usepackage{soul}
\usepackage{floatrow}
\usepackage{subfig}
\usepackage{enumitem}
\usepackage{enumerate}

\usepackage{setspace}
\usepackage[font={footnotesize}]{caption}

\theorembodyfont{\it}

\theorembodyfont{\rm}

\definecolor{cooperative}{rgb}{0.78,0.86,0.94}
\definecolor{competitive}{rgb}{0.68,0.72,0.78}
\usepackage{framed}
\usepackage{lipsum}
\usepackage{sidecap}

\def\dotquote{\begin{quote} }
\def\enddotquote{''\end{quote}}

\begin{document}

\title{Robust and tunable bursting requires slow positive feedback}

\author{Alessio Franci$^*$, Guillaume Drion$^*$, Rodolphe Sepulchre}

\date{}
\maketitle

\subsection*{Running title}
Bursting requires slow positive feedback

\subsection*{Authors and affiliations}

Alessio Franci is with Department of Mathematics,  Universidad Nacional Aut\'onoma de M\'exico, Mexico. Guillaume Drion is with the Institut Montefiore, Universit\'e de Liege, Belgium.  Rodolphe Sepulchre is with the Department of Engineering, University of Cambridge, United Kingdom. 

\subsection*{Equal contribution}

$^*$ These authors contributed equally to this work.

\subsection*{Corresponding author}

Alessio Franci, afranci@ciencias.unam.mx, tel. +5215562054874

\subsection*{Funding}

The research leading to these results has received funding from the European Research Council under the Advanced ERC Grant Agreement Switchlet n.670645 and from DGAPA-UNAM under the grant PAPIIT RA105816.

\clearpage
\doublespacing

\begin{center}
\Large
{\bf Robust and tunable bursting requires slow positive feedback}

\vspace{1cm}
{\small\noindent{\bf Abstract.}}
\end{center}
{\small\noindent We highlight that the robustness and tunability of a bursting model critically relies on currents that provide slow positive feedback to the membrane potential. Such currents have the ability of making the total conductance of the circuit {\it negative} in a time scale that is termed {\it slow} because intermediate between the {\it fast} time scale of the spike upstroke and the {\it ultraslow} time scale of even slower adaptation currents. We discuss how such currents can be assessed either in voltage-clamp experiments or in computational models. We show that, while frequent in the literature, mathematical and computational models of bursting that lack the slow negative conductance are fragile and rigid. Our results suggest that modeling the slow negative conductance of cellular models is important when studying the neuromodulation of rhythmic circuits at any broader scale.
}

\subsection*{New and noteworthy}

Nervous system functions rely on the modulation of neuronal activity between different rhythmic patterns. The mechanisms of this modulation are still poorly understood. Using computational modeling, we show the critical role of currents that provide slow negative conductance, distinct from the fast negative conductance necessary for spike generation. The significance of the slow negative conductance for neuromodulation is often overlooked, leading to computational models that are rigid and fragile.

\section*{Introduction}

While the function of neuronal bursting is still debated and probably diverse, the continuous modulation between distinct firing patterns is an important signaling component of many nervous functions. Those include muscle contraction orchestrated by central pattern generators \cite{Marder2012}, control of sleep, wakefulness and attention in thalamocortical circuits \cite{McCormick1997,Sherman2001,Bezdudnaya2006},
and sensing \cite{Krahe2004}. Voltage recordings in those references suggest robust and continuous modulations between spiking and bursting. All transitions share a sharp separation between the low frequency of spikes in tonic firing and the high frequency of spikes during bursts. They are observed across a broad range of neuronal and bursting types.

The mechanisms of this regulation are still poorly understood. At the physiological level, they seem to involve a variety of ionic currents and neuromodulators, see e.g. the review \cite{Marder2007}.  At the modeling level, most textbooks on computational and mathematical neuroscience include a chapter on bursting \cite[Chapter 9]{Izhikevich2007},\cite[Chapter 5]{Ermentrout2010a}, but the mathematical theory of bursting is based on a classification of different types of bursting according to different bifurcation mechanisms. Each bursting type is associated to a different bifurcation mechanism with little importance given to the transitions between bursting types and, most importantly, to the connection between mathematical transitions and physiological modulation.
To the best of our knowledge, a mathematical theory of bursting {\it modulation} and how it relates to physiological mechanisms has not been addressed until the recent paper \cite{Franci2014}.

In this paper we use a state-of-the-art conductance-based model, widely used in computational neuromodulation studies, to highlight a modeling feature of bursting that is critical to robustness and modulation. Specifically, mimicking the classical voltage-clamp experiment, we study the conductance of the neuron by analysing the total current response to a voltage step perturbation around the threshold potential. We aim to show that, irrespective of the modeling details, the robustness and modulation properties of the model primarily rely on the ability to modulate the slow temporal component of the conductance  from positive (in spiking mode) to negative (in bursting mode). This modulation only occurs {\it transiently} in a time scale that is {\it slow} compared to the {\it fast} time scale of the spike upstroke. The voltage-clamp signature of this slow conductance makes it model-independent and easy to  assess experimentally.

The presence of a {\it slow} negative conductance in a circuit, distinct from its {\it fast} negative conductance, is easily overlooked because of its {\it transient} nature.
We highlight the {\it transient} nature of this property with a computational experiment that only changes the kinetics of calcium channel activation from slow to fast, without affecting the balance of currents at steady-state. We show that all the modulation properties of the bursting model are lost when the calcium activation is fast, just because the slow negative conductance is no longer distinct from the fast negative conductance.

A model can exhibit a slow negative conductance only if it includes an inward current with slow activation or an outward current with slow inactivation. Such currents have been named slow regenerative in \cite{Franci2013}. By definition, a model that does not include slow regenerative currents cannot exhibit a slow negative conductance in any voltage range. We show that such models abound in the literature of bursting. This is because the slow negative conductance, while essential to robust modulation, is not necessary to bursting {\it per se}. But we illustrate on a number of published models that bursting models that lack a source of slow negative conductance are both fragile and rigid: they are very sensitive to small parameter variations, and those parameter variations disrupt the bursting pattern altogether rather than modulating the shape of the bursting pattern. In sharp contrast, published models that include a slow negative conductance are robust and tunable: small parameter variations do not disrupt the bursting pattern and specific parameter variations modulate the bursting shape between different bursting types.

The total conductance of a neuronal circuit is modulated in a given time scale by a balance between currents of negative and positive conductance, or equivalently, by a balance between currents providing positive and negative feedback to the membrane potential. Our results suggest that modulating the sign of the slow conductance of a model is necessary to the regulation of bursting, meaning that slow regenerative channels are a natural target for neuromodulators involved in bursting modulation, in line with a number of experimental studies \cite{Marder2007}.

We also provide an analysis of our results in terms of phase portraits and bifurcation theory, the classical language of bursting theory. Phase portraits of regenerative and restorative models are indeed fundamentally different \cite{Drion2012,Franci2012}. We show that only in the presence of a slow negative conductance a same phase portrait is both robust to parameter variations {\it and} compatible with various bursting types that have traditionally been associated to distinct models. This comparison suggests the relevance of analyzing bursting as circuits regulated by a balance of positive and negative feedbacks in distinct time scales as a complement to the traditional classification based on bifurcation theory.

While the analysis in this paper is performed at the single cell level, there is growing  evidence, see e.g. the recent paper \cite{dethier2015} that slow positive feedback at the cellular level critically impacts the robustness and tunability of rhythmic circuits as well. This suggests that accounting for the modeling feature highlighted in this paper is  relevant for neuromodulation studies at every scale and therefore a feature that merits attention both from experimentalists and modelers.

\section*{Methods}

{\small
All simulations and analyses were performed using the Julia programming language. The Julia code is available as Extended Data and can be downloaded at https://github.com/elsesma/eNeuro2017-Code.

Figure \ref{fig:1}A is generated using the STG model described in \cite{Goldman2001}. Briefly, the model is composed of a leak current $I_{leak}$, a transient sodium current $I_{Na}$, a T-type calcium current $I_{Ca,T}$, a S-type calcium current $I_{Ca,S}$, a delayed rectifier potassium current $I_{K,DR}$, a transient potassium current $I_{A}$, a calcium activated potassium current $I_{K,Ca}$. Parameters used in the simulations are as follows. (a): $C=1\,\mu F\cdot cm^{-2}$, $V_{Na}=50\,mV$, $V_K=-80\,mV$, $V_{Ca}=80\,mV$, $V_{leak}=-50\,mV$, $\bar g_{leak}=0.1\,mS\,cm^{-2}$, $\bar g_{Na}=700\,mS\,cm^{-2}$, $\bar g_{Ca,T}=6\,mS\,cm^{-2}$, $\bar g_{Ca,S}=9\,mS\,cm^{-2}$, $\bar g_{A}=30\,mS\,cm^{-2}$, $\bar g_{K,DR}=80\,mS\,cm^{-2}$, $\bar g_{K,Ca}=25\,mS\,cm^{-2}$. (b): same parameters as (a) except $\bar g_{Ca,T}=1\,mS\,cm^{-2}$, $\bar g_{Ca,S}=1.5\,mS\,cm^{-2}$, $\bar g_{A}=240\,mS\,cm^{-2}$. (c): same parameters as (a) except $\bar g_{Ca,T}=3\,mS\,cm^{-2}$, $\bar g_{Ca,S}=4.5\,mS\,cm^{-2}$, $\bar g_{A}=26\,mS\,cm^{-2}$. (d): same parameters as (a) except $\bar g_{Ca,T}=7\,mS\,cm^{-2}$, $\bar g_{Ca,S}=10.5\,mS\,cm^{-2}$, $\bar g_{A}=225\,mS\,cm^{-2}$. Burstiness is defined as $\frac{\text{spikes per burst}\times\text{intraburst frequency}}{\text{bursting period}}$. Voltage steps in the voltage clamp experiments are from $-40mV$ to $-39mV$.

Figure \ref{fig:1}B is generated with the same model and parameters as Figure 1A except that $\tau_{m_{Ca,T}}$ and $\tau_{m_{Ca,S}}$ are scaled by $0.5$ in the center parameter chart and $m_{Ca,T}=m_{{Ca,T}_\infty}(V),m_{Ca,T}=m_{{Ca,T}_\infty}(V)$ (instantaneous calcium activation) in the right parameter chart. Voltage clamp steps are from $-39mV$ to $-40mV$.

Nominal models in Figure~\ref{FIG:2} are given as follows. The STG model is the same as Figure~\ref{fig:1}A with maximal conductance parameters: $\bar g_{leak}=0.1\,mS\,cm^{-2}$, $\bar g_{Na}=1200\,mS\,cm^{-2}$, $\bar g_{Ca,T}=6.5.\,mS\,cm^{-2}$, $\bar g_{Ca,S}=9.75\,mS\,cm^{-2}$, $\bar g_{A}=100\,mS\,cm^{-2}$, $\bar g_{K,DR}=80\,mS\,cm^{-2}$, $\bar g_{K,Ca}=40\,mS\,cm^{-2}$. The Plant R15 model and parameters are the same as given in \cite{Rinzel1987a}. The pancreatic beta {cell} model and parameters are the same as described in \cite{Chay1983}. The thalamocortical (TC) model and parameters are the same as given in \cite{Wang1994}. The CA1 model and parameters are the same as given in \cite{Golomb2006}. The modified CA1+ model is obtained from the nominal model by: the persistent sodium current activation is made dynamic with time constant equal to $6$ times the original delayed rectifier activation time constant; the original delayed rectifier activation time constant is scaled by $4$; the cell capacitance is scaled by $0.4$.

Figure~\ref{FIG:3}A is generated using the same STG model as Figure~\ref{fig:1}A. Parameters used in the simulations are as in Figure~\ref{fig:1}A except the following. Left trace: $\bar g_{Na}=1200\,mS\,cm^{-2}$, $\bar g_{Ca,T}=1.\,mS\,cm^{-2}$, $\bar g_{Ca,S}=4.\,mS\,cm^{-2}$, $\bar g_{A}=10\,mS\,cm^{-2}$, $\bar g_{K,DR}=40\,mS\,cm^{-2}$, $\bar g_{K,Ca}=8\,mS\,cm^{-2}$. Center trace: $g_{Na}=1200\,mS\,cm^{-2}$, $\bar g_{Ca,T}=1.\,mS\,cm^{-2}$, $\bar g_{Ca,S}=7.\,mS\,cm^{-2}$, $\bar g_{A}=8\,mS\,cm^{-2}$, $\bar g_{K,DR}=40\,mS\,cm^{-2}$, $\bar g_{K,Ca}=13\,mS\,cm^{-2}$. Right trace: $g_{Na}=1200\,mS\,cm^{-2}$, $\bar g_{Ca,T}=10.\,mS\,cm^{-2}$, $\bar g_{Ca,S}=8.\,mS\,cm^{-2}$, $\bar g_{A}=10\,mS\,cm^{-2}$, $\bar g_{K,DR}=120\,mS\,cm^{-2}$, $\bar g_{K,Ca}=40\,mS\,cm^{-2}$.\\
Figure~\ref{FIG:3}B is generated using the same STG model as Figure~\ref{fig:1}A. Parameters used in the simulations are as in Figure~\ref{fig:1}A except the following. Parabolic case: $\bar g_{Na}=1200\,mS\,cm^{-2}$, $\bar g_{Ca,T}=1.\,mS\,cm^{-2}$, $\bar g_{Ca,S}=32.\,mS\,cm^{-2}$, $\bar g_{A}=40\,mS\,cm^{-2}$, $\bar g_{K,DR}=150\,mS\,cm^{-2}$, $\bar g_{K,Ca}=200\,mS\,cm^{-2}$. Square-wave case: $g_{Na}=1200\,mS\,cm^{-2}$, $\bar g_{Ca,T}=10.\,mS\,cm^{-2}$, $\bar g_{Ca,S}=8.\,mS\,cm^{-2}$, $\bar g_{A}=10\,mS\,cm^{-2}$, $\bar g_{K,DR}=120\,mS\,cm^{-2}$, $\bar g_{K,Ca}=50\,mS\,cm^{-2}$. Tapered case: $g_{Na}=1200\,mS\,cm^{-2}$, $\bar g_{Ca,T}=1.\,mS\,cm^{-2}$, $\bar g_{Ca,S}=40.\,mS\,cm^{-2}$, $\bar g_{A}=40\,mS\,cm^{-2}$, $\bar g_{K,DR}=200\,mS\,cm^{-2}$, $\bar g_{K,Ca}=200\,mS\,cm^{-2}$.

Figure~\ref{FIG:5}A,B are generated using the same STG model as Figure~\ref{fig:1}A. The bifurcation diagrams are computed by setting the calcium-activated potassium channel activation variable as the bifurcation parameter, and all other ultraslow variables at constant values ($h_{Ca,T} = h_{Ca,S} = 0.15$, $h_A = 0.05$). Bifurcation diagrams of Figure~\ref{FIG:5}A are computed using the original STG model in spiking and bursting modes. Bifurcation diagrams of Figure~\ref{FIG:5}B are computed using the STG model in bursting mode for different values of the calcium channel activation time constant. Figure~\ref{FIG:5}C is computed by simulating the STG model (left) and CA1 model (right) in bursting mode configuration for different values of the membrane capacitance ($C_m = 1\mu F/cm^2$ corresponds to the original value in both cases). Figure~\ref{FIG:5}D is computed by simulating the CA1 mode in bursting mode configuration and after changes in various model parameters as indicated in the figure. The bifurcation diagrams are computed for the original model and after an increase (top right) or a decrease (bottom right) in membrane capacitance.

Figure \ref{FIG:6}A top is generated using the STG model described in \cite{Liu1998}. Briefly, the model is composed of a leak current $I_{leak}$, a transient sodium current $I_{Na}$, a T-type calcium current $I_{Ca,T}$, a S-type calcium current $I_{Ca,S}$, a delayed rectifier potassium current $I_{K,DR}$, a transient potassium current $I_{A}$, a calcium activated potassium current $I_{K,Ca}$, and hyperpolarization-activated cyclic nucleotide-€"gated $I_{H}$ current. Parameters used in the simulations are as follow. Tonic firing: $C=1\,\mu F\cdot cm^{-2}$, $V_{Na}=50\,mV$, $V_K=-80\,mV$, $V_{Ca}=80\,mV$, $V_{leak}=-50\,mV$, $\bar g_{leak}=0.01\,mS\,cm^{-2}$, $\bar g_{Na}=800\,mS\,cm^{-2}$, $\bar g_{Ca,T}=1\,mS\,cm^{-2}$, $\bar g_{Ca,S}=1\,mS\,cm^{-2}$, $\bar g_{A}=50\,mS\,cm^{-2}$, $\bar g_{K,DR}=90\,mS\,cm^{-2}$, $\bar g_{K,Ca}=60\,mS\,cm^{-2}$, $\bar g_{H}=0.1\,mS\,cm^{-2}$. Bursting: same parameters as tonic except $\bar g_{Ca,T}=4\,mS\,cm^{-2}$, $\bar g_{Ca,S}=8\,mS\,cm^{-2}$. Voltage steps in the voltage clamp experiments are from $-44mV$ to $-42mV$.

Figure \ref{FIG:6}A bottom is generated using the same STG model as Figure 1A. Parameters used in the simulations are as in \ref{fig:1}A except $\bar g_{Na}=800\,mS\,cm^{-2}$, $\bar g_{Ca,T}=10\,mS\,cm^{-2}$, $\bar g_{Ca,S}=8\,mS\,cm^{-2}$, $\bar g_{A}=10\,mS\,cm^{-2}$, $\bar g_{K,DR}=120\,mS\,cm^{-2}$, $\bar g_{K,Ca}=50\,mS\,cm^{-2}$ (tonic mode) or $\bar g_{Na}=800\,mS\,cm^{-2}$, $\bar g_{Ca,T}=1\,mS\,cm^{-2}$, $\bar g_{Ca,S}=1\,mS\,cm^{-2}$, $\bar g_{A}=10\,mS\,cm^{-2}$, $\bar g_{K,DR}=120\,mS\,cm^{-2}$, $\bar g_{K,Ca}=50\,mS\,cm^{-2}$ (bursting mode). Voltage steps in the voltage clamp experiments are from $-44mV$ to $-42mV$.

Figure \ref{FIG:6}B bottom is generated using the Plant R15 aplysia model as described in \cite{Rinzel1987a}. Briefly, the model is composed of a leak current $I_{leak}$, a transient sodium current $I_{Na}$, a persistent calcium current $I_{Ca}$, a delayed rectifier potassium current $I_{K,DR}$, a calcium activated potassium current $I_{K,Ca}$. Parameters used in the simulation are as follows. $C=0.8\,\mu F\cdot cm^{-2}$, $V_{Na}=30\,mV$, $V_K=-75\,mV$, $V_{Ca}=140\,mV$, $V_{leak}=-40\,mV$, $\bar g_{leak}=0.003\,mS\,cm^{-2}$, $\bar g_{Na}=4\,mS\,cm^{-2}$, $\bar g_{K,DR}=4\,mS\,cm^{-2}$, $\bar g_{Ca}=0.006\,mS\,cm^{-2}$, $\bar g_{K,Ca}=0.04\,mS\,cm^{-2}$. Voltage steps in the voltage clamp experiments are from $-80mV$ to $-40mV$. Figure \ref{FIG:6}B top is generated using the same model and parameters as Figure \ref{FIG:6}B bottom, except that the calcium current activation is 100 times faster.

}

\section*{Results}

\subsection*{A transient signature of robust and tunable bursting}
\label{sec:2}

\begin{figure}
\centering
\includegraphics[width=0.9\textwidth]{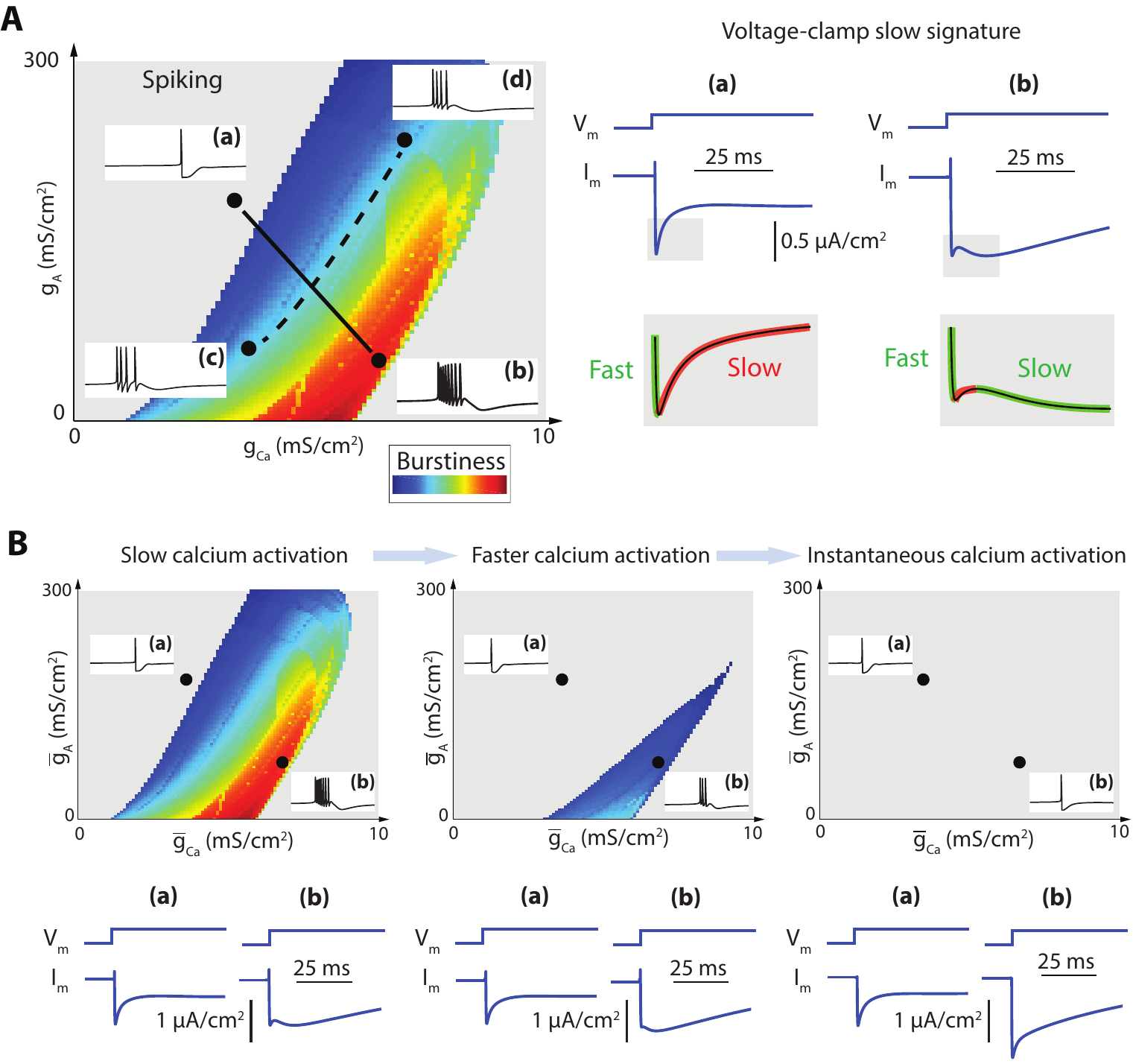}
\caption{\protect A. The slow transient in a voltage clamp experiment near threshold is a reliable  signature to discriminate between bursting and tonic firing. Left panel: parameter chart of burstiness as a function of the parameters $g_A$ and $g_Ca$ in the STG model \cite{Goldman2001}. Burstiness was computed as described in the Method Section. Right panel: voltage clamp experiment close to threshold potential in tonic mode (a) and bursting mode (b). {Voltage steps are from $-40mV$ to $-39mV$.} The slow transient of the {current response to a voltage step}  is increasing in spiking mode (a signature of slow positive conductance) and decreasing in bursting mode (a signature of slow negative conductance). The signature is modulated along a modulation path (a-b) and conserved along  a degeneracy  path (c-d, not shown) in the parameter space of maximal conductances. B. Effects of decreasing calcium current activation time constant on the parameter chart and the voltage clamp experiment in A. The decreasing phase of the slow transient vanishes as calcium activation kinetics, the only source of slow negative conductance in the model, varies from slow to fast. In the parameter charts, reduction of the calcium activation time constants shrinks the parameter region where the model can be modulated. In the limit of instantaneous activation, the model has lost its modulation properties and in particular the transition from tonic firing to bursting.
}\label{fig:1}
\end{figure}

Figure~\ref{fig:1}A uses the computational model of \cite{Goldman2001} to illustrate a classical physiological transition from tonic firing to bursting. The  model includes seven voltage- and time-dependent conductances as well as a leak conductance (see Methods). The modulation from tonic firing to bursting is obtained by varying the balance of calcium and A-type potassium voltage-gated currents. This modulation defines modulatory paths in the parameter space of the two maximal conductances $\bar{g}_{Ca}$ and $\bar{g}_A$ (Path a-b in Figure \ref{fig:1}A ). The same plane contains degeneracy paths where modulation of the maximal conductances results in almost no change in the resulting neuronal activity (Path c-d in Figure \ref{fig:1}A ). The coexistence of degeneracy and modulatory paths has been shown to be critical for robust neuromodulation \cite{Marder2014}.  A computational model that reproduces such features does not only exhibit a bursting trace for a well chosen set of conductance parameters. In addition, the bursting rhythm is robust and tunable in the parameter space of maximal conductances. 

Figure \ref{fig:1}B highlights that this tunability property is completely lost by changing a single parameter in the model, namely, the time constant of activation of the calcium channels.
Before interpreting this result, we stress that the modeling difference between Figure \ref{fig:1}A and Figure \ref{fig:1}B is purely {\it dynamical} in nature: it does not affect the {\it static} behavior of the model, that is, the model equations at equilibrium. This means in particular an identical balance of ionic currents at equilibrium and an identical I-V curve.

To unfold the {\it transient} mechanism responsible for the structural change between Figure \ref{fig:1}A and Figure \ref{fig:1}B, we mimick the classical voltage-clamp experiment of electrophysiology: we clamp the voltage at a constant value close to threshold potential ($V_{th}\sim-40mV$ in this model) and apply a small voltage step perturbation $\Delta V_m$ at time $t=0$.  The current step response $\Delta I_m(t)$ provides us with the temporal evolution of the local {\it conductance} $\Delta I_m(t)/ \Delta V_m $ of the model around the threshold potential. This total conductance is the aggregate conductance resulting from all the ionic current variations at a given time and around a given voltage.

In Figure \ref{fig:1}A, we see that the {\it transient} behavior of the local conductance is markedly different in the spiking configuration (a) and in the bursting configuration (b). In the spiking configuration, the current step response carries the usual signature of an excitable circuit: an initial phase characterised by a fast inverse response, followed by a slow monotone convergence to equilibrium. In the bursting configuration, the current response exhibits an additional slow inverse response, distinct from the initial fast inverse response. If we decompose the current response into  fast, slow, and ultraslow transient phases, it is the slow transient that discriminates bursting from spiking. 

In Figure \ref{fig:1}B, the distinct slow transient signature of bursting is progressively lost as the time constant of calcium activation is decreased. This is because the two distinct inverse responses  progressively merge. In the limiting case of an instantaneous calcium activation, they simply add up in the fast time scale. This phenomenon is easy to explain in the computational model used in Figure \ref{fig:1}: the first ({\it fast}) inverse response of the current results from the fast activation of sodium channels whereas the second ({\it slow}) inverse response results from the slow activation of calcium channels. The two successive inverse responses are distinct in the voltage clamp experiment of Figure \ref{fig:1}A  because the time scales of the calcium channel activations are significantly slower than the time scale of the sodium channel activation \cite{Kostyuk1977}, \cite[p.127]{Hille2001}. In contrast, the two successive inverse responses merge in Figure \ref{fig:1}B because the time scale of calcium channel activation merges with the time scale of sodium channel activation.

\subsection*{A robust and tunable bursting model must include a source of slow negative conductance}
\label{sec:3}

In the seminal work of Hodgkin and Huxley, the voltage clamp experiment described in the previous section was applied to the squid giant axon and served as a foundation to model the voltage dependence and the dynamics of ionic conductances. The {\it early inverse} current response was attributed to a {\it fast negative} conductance modeled by the {\it fast} activation of an inward (sodium) current. Likewise, the {\it late monotone} convergence was attributed to a {\it slow positive} conductance modeled both by the {\it slow} inactivation of sodium channels and the {\it slow} activation of an outward (potassium) current.

We proceed in the same manner to explain the slow transient signature of a bursting neuron: it requires a {voltage-gated current} providing a {\it slow negative conductance}, distinct from the {\it fast} negative conductance. A negative conductance is provided by the activation of an inward current or the inactivation of an outward current. It is {\it slow} if the corresponding channel kinetics is distinctively slower than the fast activation of sodium channels {and distinctively faster than adaptation current kinetics. Typical slow conductance time constants are in the range $5-20\,{\rm ms}$}. Physiological contributors of such currents include the whole family of calcium currents with slow activation as well as resurgent sodium channels \cite{Swensen2003}. They also include any outward current that inactivates slowly, {such as some potassium channels} \cite{Storm1990}. Such channels have been named {\it slow regenerative} in the paper \cite{Franci2013}. 

Our computational experiment suggests that a robust and tunable bursting neuronal model must include a source of slow negative conductance. For a conductance-based model, this means that the gated ionic currents must include at least one type of slow regenerative channel. In the simulated STG model, only calcium currents contribute to the slow negative conductance. They do so  because their activation is slow. The modulation path in Figure \ref{fig:1}A amplifies the slow negative conductance of the total current from (a) to (b) by modulating the balance between slow regenerative (calcium) channels and slow restorative (potassium) channels. In Figure \ref{fig:1}B, this modulation property is lost because the calcium channels become fast regenerative. Modulation of the total conductance in the slow time scale from positive to negative is no longer possible because the model has lost its only source of slow regenerative channels.

In a conductance-based model, modulation of the total slow negative conductance is possible only in the presence of slow regenerative channels. The voltage-clamp experiment in the previous section is a general method to assess the negative slow conductance of a circuit, irrespective of the modeling details of the model. The reader is referred to the recent paper \cite{Drion2015} for a method that quantitatively assesses the slow negative conductance (or any other conductance) of an arbitrary one-port circuit at a given voltage, either computationally or experimentally. 

\subsection*{Robust versus fragile bursting}
\begin{figure}
\centering
\includegraphics[width=0.9\columnwidth]{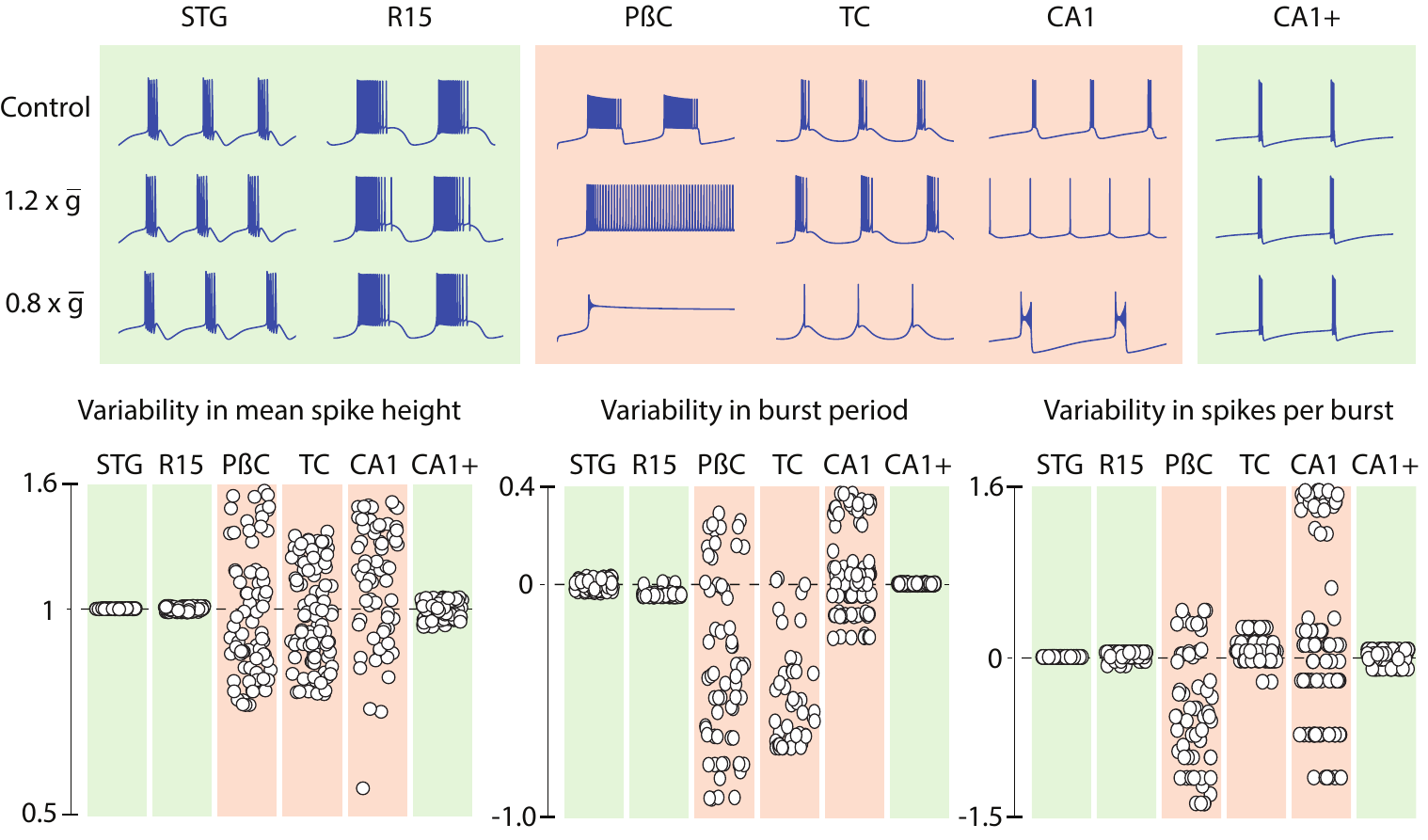}
\caption{Bursting models that lack currents providing slow negative conductance are fragile: tiny parameter variations   disrupt the nominal rhythm. Green models (STG and R15) do include slow regenerative channels providing slow negative conductance. Red models (p$\beta$C,  TC, and CA1) lack slow negative conductance. {\it Top panels}: only the bursting traces of green models are robust to a uniform scale of the maximal conductance vector ($\bar{\mathbf{g}}\mapsto\,0.8\bar{\mathbf{g}}$ or $\bar{\mathbf{g}}\mapsto\,1.2\bar{\mathbf{g}}$). The red model CA1 is turned into the robust green model CA1+ by making the calcium activation slow. See Methods for details. {\it Bottom panels}: random uniform scaling of the vector of maximal conductances induce large variability in the rhythm properties only in models lacking slow negative conductance. The scatter plots are obtained by scaling the maximal condutance vector $\bar{\mathbf{g}}$ by a uniformly distributed random number in the range $[0.8,1.2]$. Variability plots are absolute for the mean spike height (left) and logarithmic for the burst period (center) and number of spikes per burst (right): $\frac{\text{(spike height)}}{\text{(spike height)}_{\rm nominal}}$, $\log\left(\frac{\text{(burst period)}}{\text{(burst period)}_{\rm nominal}}\right)$, $\log\left(\frac{\text{(spikes-per-burst)}}{\text{(spikes-per-burst)}_{\rm nominal}}\right)$.}\label{FIG:2}
\end{figure}

The absence of slow negative conductance has a dramatic consequence on the robustness of the bursting model to parameter perturbations. Figure \ref{FIG:2} illustrates the striking contrast between the fragility of models that lack slow negative conductance and the robustness of models that include slow negative conductance. The chosen perturbation is a uniform scaling of all maximal conductance parameters, which is mathematically equivalent to a scaling of the membrane capacitance.

The STG \cite{Goldman2001} and the R15 \cite{Rinzel1987a} models are two classical bursting models of the literature that are robust to the perturbation. Both include slow regenerative channels by modeling calcium channels that activate {\it slowly}. The three models p$\beta$C \cite{Chay1983},  TC \cite{Wang1994}, and CA1 \cite{Golomb2006} are three published models that are fragile to the perturbation. Small deviations from the nominal parameter set produce large variations in different properties of the rhythm. The three models lack any source of slow negative conductance. They all include calcium channels or other regenerative channels but assume an instantaneous activation, making them {\it fast} regenerative instead of {\it slow} regenerative.  

The model CA1+ is a modification of the published CA1 model. In the modified model, the activation of persistent sodium channels was modified from fast to slow, making the persistent sodium channels slow regenerative instead of fast regenerative (see Methods for details). This only modification was sufficient to recover the robustness of models that have a slow negative conductance.

\subsection*{Tunable versus rigid bursting}
\begin{figure*}
\centering
\includegraphics[width=0.9\textwidth]{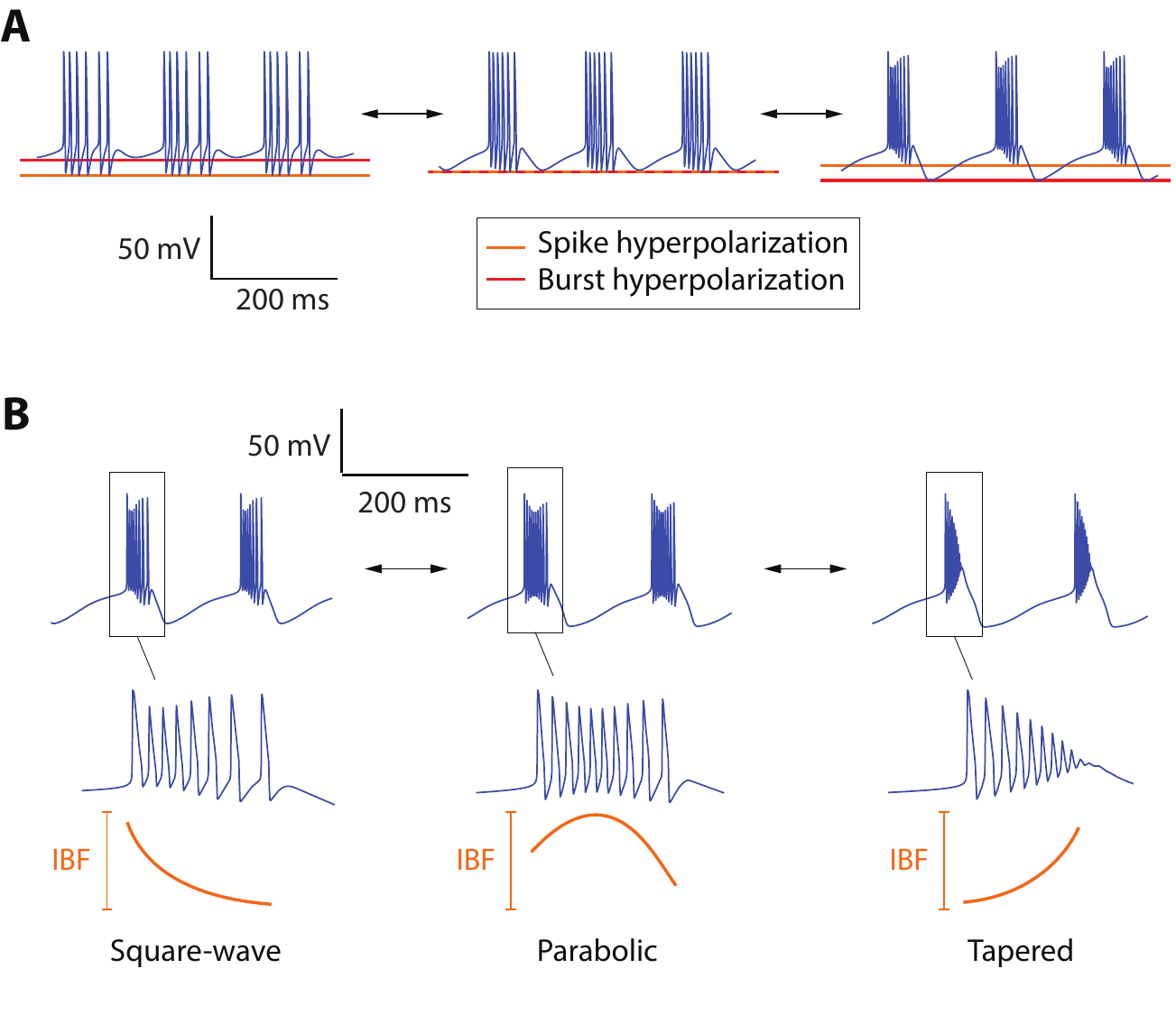}
\caption{Bursting models that include currents providing slow negative conductance are tunable. A. The bursting quality (intraburst frequency, plateau or non-plateau) can continuously be modulated via changes in ion channel densities independently of the bursting rhythm (interburst frequency). B. The bursting shape can be continuously modulated via changes in ion channel densities while maintaining both intraburst and interburst frequencies (IBF = intraburst frequency). See methods for the used parameter values.}\label{FIG:3}
\end{figure*}

Bursting models that include a slow negative conductance are not only robust, but they are robustly tunable. This means that the shape of the bursting trace can be tightly controlled by modulating maximal conductance values (i.e. channel densities). Figure \ref{FIG:3} illustrates those modulation properties with the same published STG model as in the previous sections. Figure \ref{FIG:3}A illustrates the modulation of bursting quality: the intraburst frequency and plateau properties are continuously modulated while maintaining other features of the burst unaffected, such as the interburst frequency or the mean voltage during resting and spiking phases. Figure \ref{FIG:3}B illustrates the modulation of the bursting shape, while maintaining the mean intraburst and interburst frequencies. The continuous modulation recovers three distinct types of bursting usually referred to as ``square-wave", ``parabolic", and ``tapered".

The modulation properties illustrated in Figure \ref{FIG:3} do not result from a systematic exploration of the parameter space, a task already formidable for the chosen STG model \cite{Prinz2003}. Instead, they only rely on modulating the ratio of the maximal values of the total slow negative conductance and of the total ultra-slow positive conductance following the methodology of dynamic input conductances \cite{Drion2015}. Shaping the dynamic conductances relatively to each other is easy and intuitive because the tuning parameters are few and directly map to the bursting behavior. This qualitative tuning is then easily translated into physiologically plausible modulations. For instance, modulation of the slow negative conductance relative to the ultraslow positive conductance in Figure \ref{FIG:3} was achieved by modifying only the five following maximal conductances: $\bar{g}_{Ca,T}, \bar{g}_{Ca,S}, \bar{g}_{A}, \bar{g}_{Kd}, \bar{g}_{KCa}$.  In each case, the modulation is robust, that is, not sensitive to small parameter variations in the large-dimensional space of the conductance-based model parameters. 

The robust modulation illustrated in Figure \ref{FIG:3} is in sharp contrast with the rigidity of bursting models that lack a source of slow negative conductance. The nominal bursting trace 
of the three models p$\beta$C,  TC, and CA1 in Figure \ref{FIG:2} is rigid because the relationship between intraburst and interburst frequencies as well as the relationship between the mean voltage of the resting and spiking modes is extremely constrained. The resulting burst is not only fragile; it is also rigid, making it difficult to modulate the burstiness or the bursting type as in Figure \ref{FIG:3}. The geometric analysis in the next section provides additional insight to those limitations.

\subsection*{Connection with phase portrait and bifurcation analysis}
The critical role of the slow negative conductance for modulation and robustness will now be examined in the light of geometric analysis. We rely on the common simplification that a three time-scale bursting attractor can be analyzed via the bifurcation diagram of a two time-scale phase portrait. The variables of the phase portrait are the fast voltage and a slow variable aggregating all the slow variables. The bifurcation parameter is a representative ultraslow variable. Bursting is modeled as ultraslow adaptation of the bifurcation parameter across a parameter range where a stable fixed point (the {\it resting} state) and a stable limit cycle (the {\it spiking} state) coexist in the phase portraits.

\begin{figure}
\centering
\includegraphics[width=0.9\textwidth]{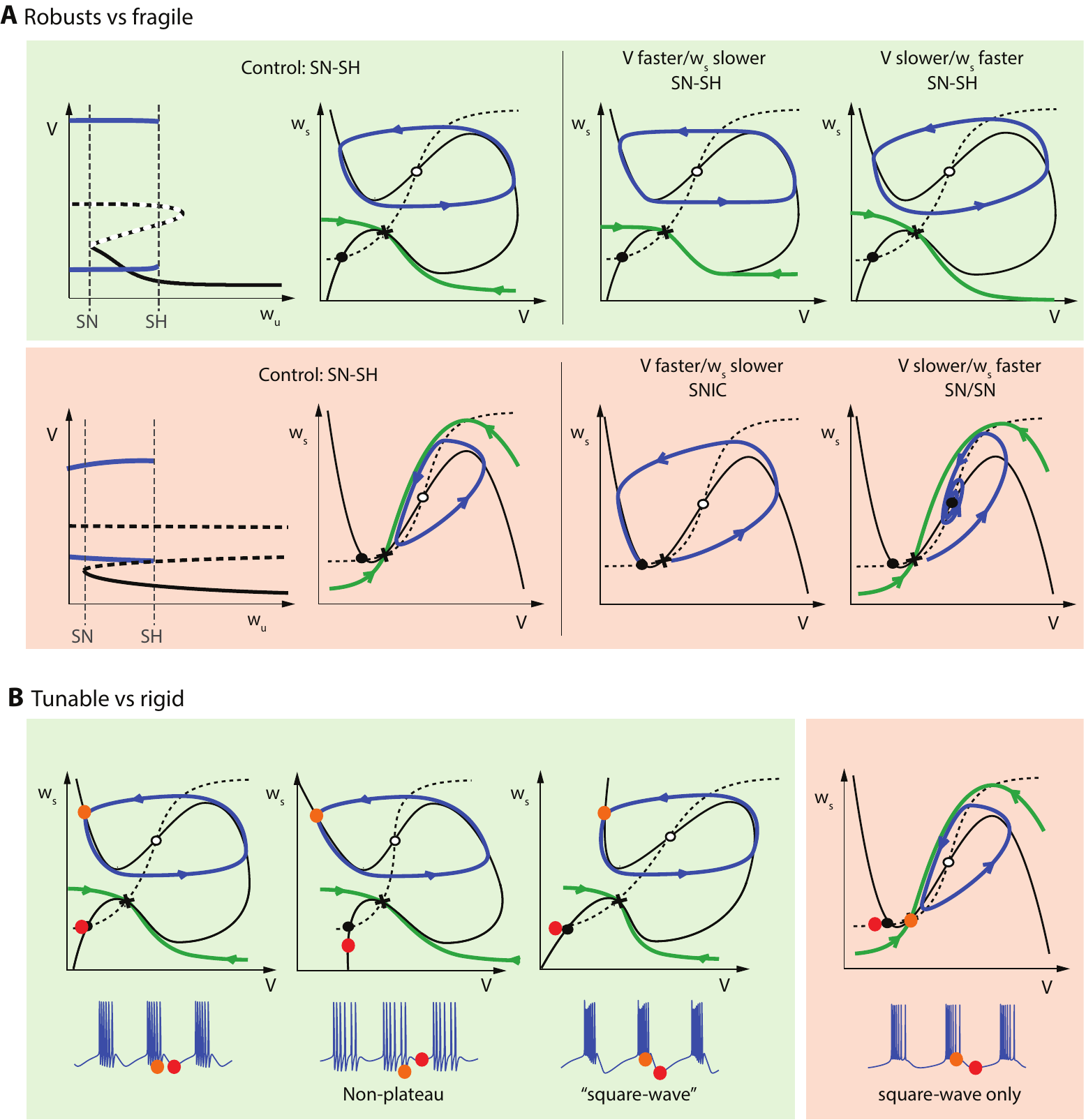}
\caption{The phase portrait geometry of regenerative and restorative bursting. A. While sharing the same saddle node (SN) and saddle-homoclinic (SH) bifurcations, the phase portraits of rest-spike bistable models are radically affected by the slow negative conductance. Regenerative phase portraits (green) are robust to variations of the fast and slow time scales. Restorative phase portraits (red) are rest-spike bistable only for a well-chosen ratio of time scales. Stable fixed points are marked as full dots, unstable fixed points as empty dots, saddle points as crosses. Limit cycles and typical trajectories are sketched as blue oriented lines. The stable manifold of saddle points is sketched as a green oriented line. Bifurcation diagrams are the same as Figure~\ref{FIG:5}.A bottom (regenerative case) and Figure~\ref{FIG:5}.D left (restorative case).
The mirrored N-shaped nullcline of the regenerative case ensures robustness with respect to variations in the time scale separation between the membrane potential and the recovery variable. The N-shaped nullcline of the restorative case makes the phase portrait fragile to the ratio of fast and slow time scales. Small deviations from the nominal ratio destroy the rest-spike bistability, either via a saddle-node homocinic bifurcation (cf. \cite[Figure 6.44]{Izhikevich2007}) or via Hopf bifurcation around the up equilibrium. B. Regenerative phase portraits can be continuously deformed to match different types of bursters. Only the square-wave burster is compatible with restorative phase portraits.}\label{FIG:4}
\end{figure}

The role of the slow negative conductance has been previously analyzed in fast-slow phase portraits (see \cite{Drion2012},\cite{Franci2012},\cite{Franci2013}) and in mathematical three-time scale models of bursting \cite{Franci2014}. The results of this qualitative analysis are summarized in Figure \ref{FIG:4}. The reduced phase portraits are called regenerative when they include a slow negative conductance and restorative otherwise. Figure \ref{FIG:4}A shows that the two types of phase portrait are qualitatively different. In restorative phase portraits, the $V$-nullcline has the classical  $N$-shape of spiking neuronal models (red). In regenerative phase portraits, this shape is mirrored (green). The reader is referred to \cite{Drion2012},\cite{Franci2012},\cite{Franci2013} for a detailed analysis of why the mirrored $V$-nullcline requires a slow negative conductance. Both restorative and regenerative phase portraits  exhibit bistability between a fixed point and a limit cycle and both phase portraits rely on the same bifurcations: they are of the same  saddle-node/saddle homoclinic type according to the classification of \cite[Figure~9.24]{Izhikevich2007}. However,  the difference in their $V$-nullclines strongly affects the robustness and the tunability of the bistable attractor. In the regenerative phase portrait (green), the stable manifold of the saddle point (dark green) is a separatrix that sharply divides the phase portrait into two distinct regions, each of which corresponds to the basin of attraction of one of the two attractors: the stable fixed point and the stable limit cycle. The two stable attractors can be shaped independently from each other by deforming the nullclines away from the separatrix and modulating the ratio of the fast and slow time scales. Robust bistability is maintained across a broad range of variations (Figure \ref{FIG:4}A, top right). In sharp contrast, the bistability of a restorative phase portrait (green) requires both a specific intersection of the nullclines and a specific ratio between the fast and slow time scales. The bistability is fragile to any perturbation of this specific tuning (Figure \ref{FIG:4}A, bottom right). As a result, the bifurcation diagram associated to bursting only exists in a narrow parameter range.

Figure \ref{FIG:4}B illustrates how the geometry of the bistable phase portrait impacts not only the robustness but also the tunability of the bursting attractor. The resting and spiking attractors can be shaped independently in a regenerative phase portrait because deforming the $V$-nullcline near the fixed point does not affect the limit cycle and vice versa.
As a result, the values of the membrane potential at rest and in spiking mode can be tuned independently, leading to the generation of both non plateau bursting and plateau (or square-wave) bursting depending on the maximal conductance parameter set (Figure \ref{FIG:4}B, left).
This flexibility does not exist in restorative phase portraits. In particular, the resting state is always more hyperpolarized than the spiking state, forcing a bursting trace of the square-wave type (Figure \ref{FIG:4}B, right).

\begin{figure}
\centering
\includegraphics[width=0.9\textwidth]{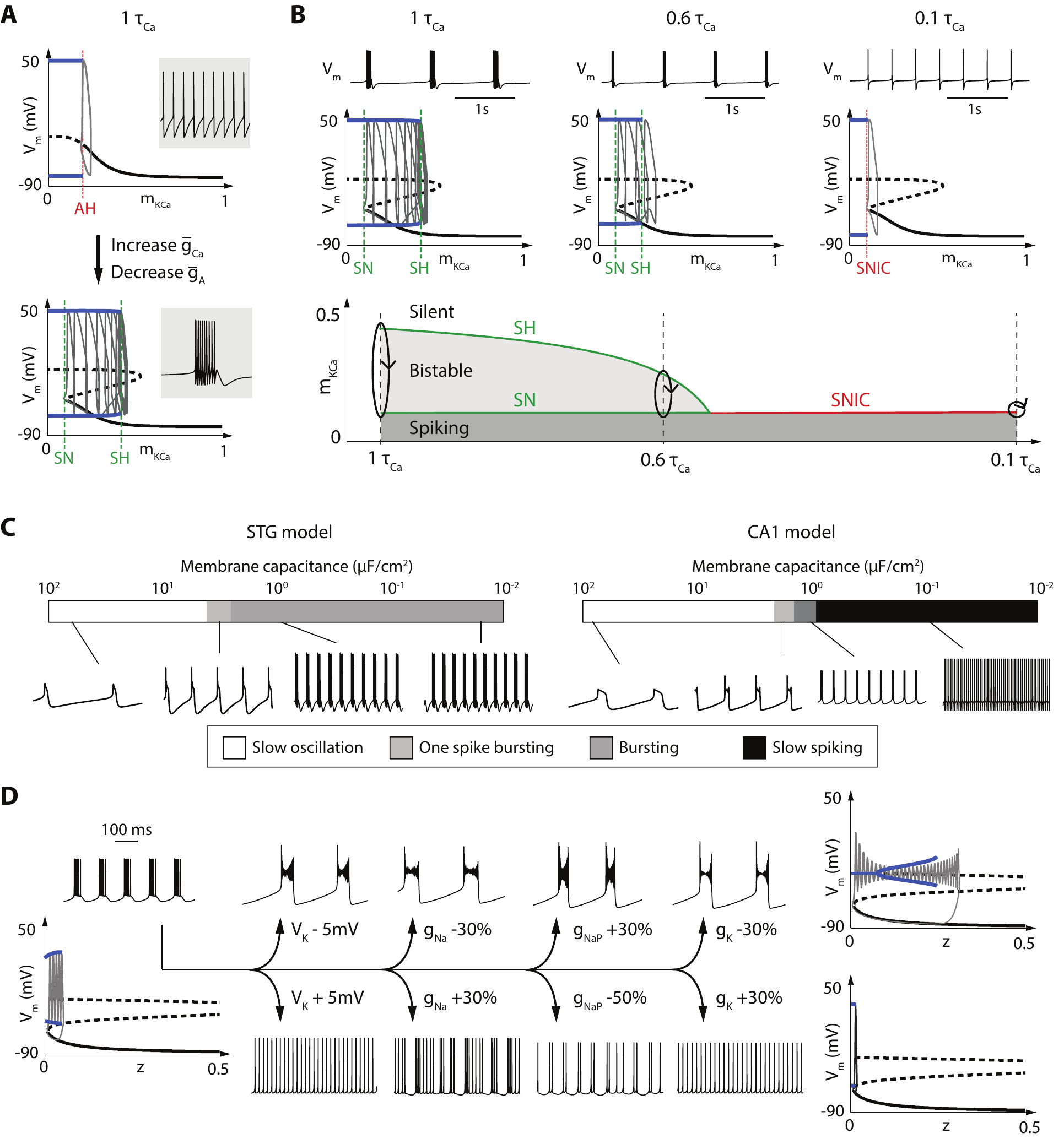}
\caption{Bifurcation analysis of bursting with and without slow negative conductance. A. Membrane potential-ultraslow  adaptation variable ($m_{K,Ca}$) bifurcation diagram and associated trajectories in tonic (top) and burst (bottom) modes in the STG model of Figure~\ref{fig:1}. Other ultra-slow variables of the model ($h_{Ca,T}$, $h_{Ca,S}$ and $h_A$) are fixed at physiologically plausible constant values (see Methods). B. Deformations of the bursting traces (top) and bifurcation diagram (center) of the STG  model with parameter as in A bottom under reduction of the calcium activation time constants. The parameter range of bistability gradually shrinks as the time constant of calcium activation decreases to zero (bottom). C. Effects of varying the cell membrane capacitance in models with (left) and without (right) slow negative conductance. In the STG model, bursting persists for arbitrary small values of the capacitance. In the CA1 model, bursting only persists in a tiny window around the nominal value of the published model. D. The fragility of bursting in CA1 model is illustrated with respect to different parameters. The different perturbations cause similar alterations of the bifurcation diagram and of the corresponding rhythm: reduction or elimination of spiking in one direction, reduction or elimination of the slow bursting oscillation in the other direction. Bifurcation diagrams correspond to: nominal case (left), increased membrane capacitance (top right), decreased membrane capacitance (bottom right).}\label{FIG:5}
\end{figure}

The qualitative analysis above is based on a low-dimensional mathematical model but the conclusions persist in higher-dimensional computational models. Figure~\ref{FIG:5} illustrates the various bifurcation diagrams associated to the numerical observations in the previous sections.

Fig~\ref{FIG:5}A  illustrates how the modulation from tonic spiking to bursting in Figure~\ref{fig:1}A  affects the corresponding bifurcation diagrams. For the tonic spiking mode (point (a) in Figure~\ref{fig:1}A), the bifurcation diagram is the bifurcation diagram of a spiking model: the equilibrium curve is monotone, and there is no bistable range. For the bursting mode (point (b) in Figure~\ref{fig:1}A), the diagram instead exhibits the bistable range of a robust saddle-homoclinic burster consistent with the regenerative phase portraits of Figure \ref{FIG:4}A.  As illustrated in Figure~\ref{FIG:5}B, the bistable range of the burster is progressively lost when the slow conductance becomes fast. Here the bistability is lost not because of a deformation of the equilibrium curve (consistent with  the fact that the {\it static} properties of the model are unchanged) but because the two bifurcations that determine the parameter range of bistability (saddle-node and saddle-homoclinic) progressively merge to a saddle-node homoclinic (SNH) bifurcation. Near the SNH bifurcation, the fragility of the bistable range is consistent with the fragility of restorative phase portraits of Figure \ref{FIG:4}A.

The impact of the bistable parameter range on the fragility of bursting is further illustrated in Figure~\ref{FIG:5}C, where we reexamine the robustness of bursting models to a perturbation of the membrane capacitance (see Figure~\ref{FIG:2}). This perturbation affects the time-scale separation between the fast and slow variables of reduced phase portraits. Bursting in the STG model is robust to this perturbation as far as the capacitance is low enough to allow for spike generation, consistently with the regenerative phase portraits of Figure \ref{FIG:4}A. The time-scale separation between fast and slow variables can be increased at will without destroying the bistable bifurcation parameter range that is essential to bursting (Figure~\ref{FIG:5}C, left). In sharp contrast, bursting in the CA1 model is fragile to the same perturbation, consistently with the restorative phase portraits of Figure \ref{FIG:4}A. The bistable parameter range quickly disappears, perturbing the SN-SH bifurcation diagram of a bursting attractor into one of two possible scenarios: either the SNIC bifurcation diagram of a slow spiker (smaller values of membrane capacitance) or the SN-SN bifurcation diagram of a slow rhythm that switches between a low and high resting states (larger values of membrane capacitance) (Figure~\ref{FIG:5}C, right). In the case where the capacitance is low enough to allow for spike generation, the robust firing pattern of the CA1 model is a slow spiking pattern, not a bursting one.

Finally, Figure~\ref{FIG:5}D illustrates that this fragility is generic and not specific to a particular parameter. The robustness of the nominal CA1 bursting model was tested against five different parameter perturbations. In each case, the perturbation is well in the range of physiological variability and it produces the same alteration of the bifurcation diagram (the effect of a $5$ mV change in the potassium reversal potential is particularly striking). The bursting attractor is {\it fragile}, in the sense that small parameter variations disrupt the rhythm of the nominal model. It is also {\it rigid}, in the sense that different parameter variations always disrupt the rhythm in the same way (i.e. deform the bifurcation diagram in the same way), disrupting either the fast or the slow oscillation of the bursting rhythm.



\section*{Discussion}

\subsection*{A transient signature characterizes the transition from spiking to bursting}
The computational experiments in this paper suggest the critical role of the {\it slow} conductance in a robust and tunable bursting model. The classical voltage clamp experiment of electrophysiology near threshold provides a clear signature that the sign of this conductance is modulated during a continuous modulation from spiking to bursting. We now discuss several reasons why the role of the slow negative conductance is often overlooked both in experimental and modeling studies.

The specific signature of a slow negative conductance has been reported experimentally, at least in two published papers: \cite[Figures~3 and~5]{Rodriguez2013} use this signature to assess the modulatory effects of the peptide CabPK in regulating a cellular and circuit bursting rhythm and  \cite[Figure 11]{Swensen2005} uses the same signature to assess the cooperation of calcium currents and persistent sodium currents in Purkinje cell burst excitability. One important reason why this signature is not more prevalent in experimental studies is that most voltage-clamp experiments nowadays study the current response to a quasi-static voltage {\it ramp} rather than a voltage {\it step}. There are several reasons to prefer a ramp to a step in an experimental setup. For instance, a single ramp input experiment might be sufficient to extract the entire {\it static} I-V curve. Unfortunately, a ramp input will mask the transient signature reported in this paper. The voltage-clamp experiments in \cite{Rodriguez2013,Swensen2005} are rare instances of {\it step} input experiments.

\begin{figure}
\centering
\includegraphics[width=0.75\textwidth]{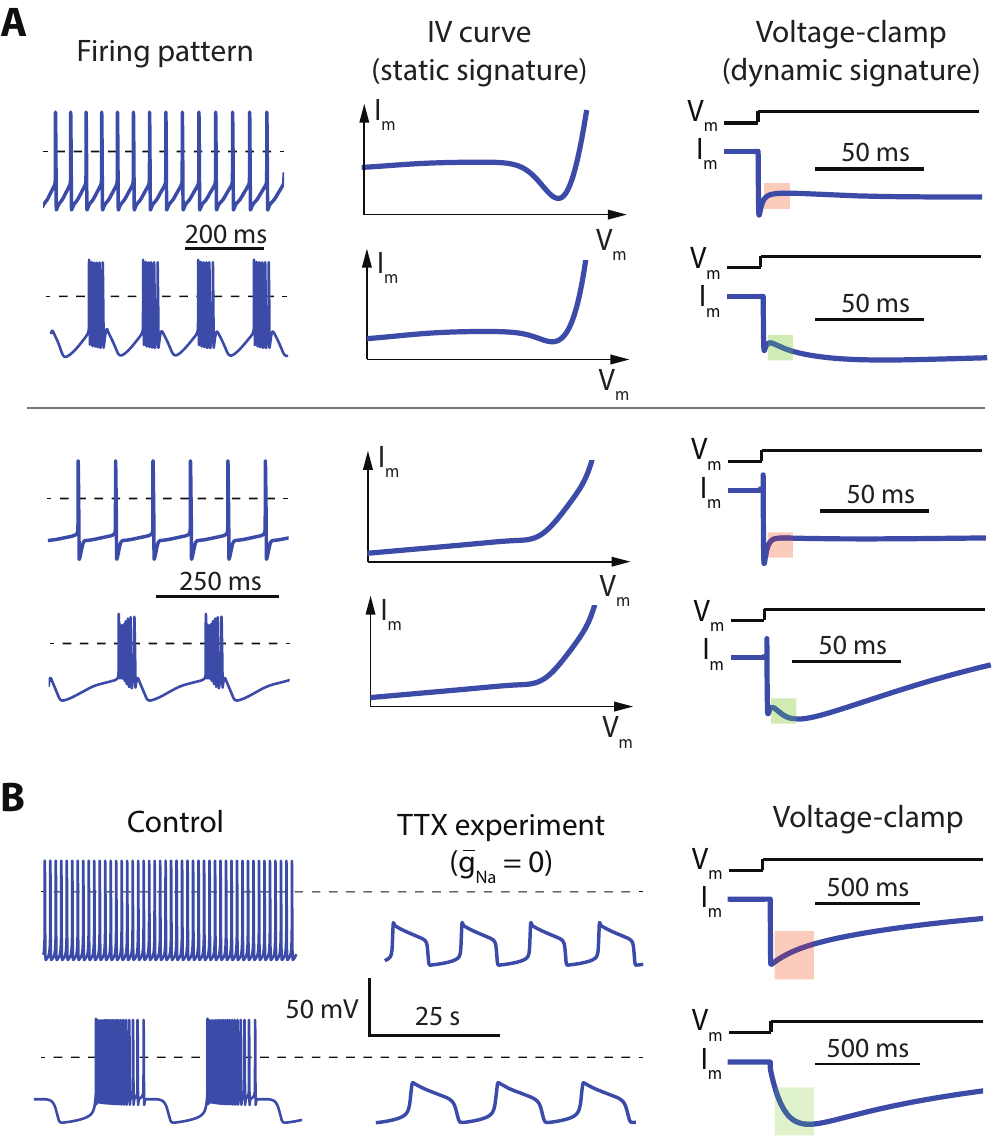}
\caption{A. The transient signature of slow negative conductance cannot be recovered from the static I-V curve. A. A region of negative conductance in the I-V curve is neither necessary nor sufficient for bursting. In the STG model described in \cite{Liu1998} model (top) the I-V curve possesses a region of negative conductance both in tonic and burst modes. In the STG model of  \cite{Goldman2001}, the I-V curve is monotone both in tonic and burst modes.  See Methods for details about the simulations.
B. Slow oscillatory potentials in the absence of sodium do not necessarily discriminate between fast and slow negative conductances. The Plant R15 aplysia model described in \cite{Rinzel1987a} exhibits slow oscillatory potential under blockade of sodium channels both in tonic (top) and burst (bottom) modes. See Methods for details about the simulations.}\label{FIG:6}
\end{figure}

More generally, it is the {\it transient} nature of the slow conductance that makes it difficult to assess from the {\it stationary} indicators traditionally associated to experimental investigations of bursting. Most importantly, such indicators include the monotonicity of the I-V curve \cite{Wilson1974}, \cite{Chen2000}, \cite{Butera2005}, \cite{Lewis1984} or the analysis of slow oscillatory potentials (SOPs) \cite{Amini1999}, \cite{Canavier2007}, \cite{Zhang2013}, \cite{Wang1992}, \cite{Wang1993a}, \cite{Skinner1994}. Figure \ref{FIG:6}A contrasts the dynamical voltage clamp signature with the monotonicity properties of the static I-V curve. It provides four model conditions for which the electrophysiological distinction between tonic firing and bursting is unambiguously predicted by the slow transient voltage clamp signature but not by the $I-V$ curve. This is because the I-V curve only depends on the stationary value of a voltage-clamp experiment, but not on the {\it transient} current response. By definition, the slope of the I-V curve at a given voltage is the local conductance of the model at steady-state: for a step of small amplitude, this corresponds to the asymptotic value for large $t$ of the ratio $\Delta I_m(t) / \Delta V_m$. A negative conductance in the I-V curve is thus synonym of an inverse {\it steady-state} response in the voltage clamp experiment. The slow transient of the voltage clamp is a {\it transient} feature that cannot be inferred from the steady-state response of the voltage clamp.

Bursting and its modulation are also often studied experimentally through slow oscillatory potentials (SOPs) observed in the absence of spikes (by blocking sodium channels). But a slow oscillation is not a reliable signature of bursting per se either. Fig \ref{FIG:6}B illustrates that a slow oscillation does not necessarily discriminate between tonic firing and bursting because either a fast or a slow current can be responsible for the regenerative upstroke of the slow oscillation. Fast or instantaneous activation of a calcium channel as in Figure \ref{fig:1}B will generate neither the slow transient voltage clamp signature nor bursting. But it provides a steady-state inward current that can be sufficient to destabilize the resting potential and generate a slow oscillatory potential in the absence of sodium channels. Once again, the dynamic role of a given current cannot be inferred from its static properties.

The negative slow conductance is also often overlooked in modeling and computational studies. Its role can only be captured in models that respect the time scale separation between fast and slow regenerative channels. This time scale separation is well acknowledged in the ion channel literature. For instance, activation and inactivation of calcium channels is often described as similar to activation and inactivation of sodium channels, but up to fifty times {\it slower} for some of them \cite{Kostyuk1977}, \cite[p.127]{Hille2001}. But it is often neglected in mathematical and computational modeling. For instance, Figure 5.6 in the textbook \cite{Izhikevich2007} refers to both sodium and calcium activation as {\it fast}. The section on calcium channels in the recent textbook \cite{Ermentrout2010a} also suggests that calcium and sodium channels have similar dynamics. In computational modeling, it is widespread practice to set both the calcium and sodium activation to steady-state when reducing the complexity of a  model. See Table~\ref{TAB1} for a list of important papers that make that assumption. All those references suggest that the role of slow negative conductances in robustness and neuromodulation is underappreciated. 

\begin{table}[h!]
\centering
\begin{tabular}{|r|l|}
\hline
{\bf Reference} & {\bf Slow regenerative gating variable set to steady state}\\
\hline
\cite{Terman2002} & Activation of T-type and high-threshold $Ca^{2+}$ channels\\
\cite{Rubin2004} & Activation of T-type $Ca^{2+}$ channels\\
\cite{Butera1999} & Activation of persistent $Na^+$ channels\\
\cite{Butera1999a} & Activation of persistent $Na^+$ channels\\
\cite{Pospischil2008} & Activation of T-type $Ca^{2+}$ channels\\
\cite{Rush1994} & Activation of T-type $Ca^{2+}$ channels\\
\cite{Smith2000} & Activation of T-type $Ca^{2+}$ channels\\
\cite{Kubota2011} & Activation of T-type and high-threshold $Ca^{2+}$ channels\\
\cite{Golomb1997} & Activation of persistent $Na^+$ channels\\
\cite{Wang1994} & Activation of T-type $Ca^{2+}$ channels\\
\cite{Golomb2006} & Activation of persistent $Na^+$ channels\\
\hline
\end{tabular}
\caption{List of {published models lacking a slow negative conductance because the activation of  slow regenerative channels is set to steady-sate.}}\label{TAB1}
\end{table}

%

\subsection*{Classification versus modulation of bursting models}
Mathematical models of bursting often omit the slow negative conductance because they only include the minimal number of currents that is necessary to bursting. The rationale is simple: a spiking model only requires two distinct currents to model the fast negative and the slow positive conductances. In such a model, the modulation from rest to spike is achieved by modulating the constant applied current. A third current is then enough to model the additional ultraslow positive conductance that converts the spiker into a burster. In this approach, bursting is seen as the result of ultraslow adaptation between resting and spiking. This minimal motif only requires three distinct ionic currents and is at the core of textbook expositions of bursting such as Chapter 9 in \cite{Keener2009}, Chapter 9 in \cite{Izhikevich2007}, and Chapter 5 in \cite{Ermentrout2010a}. It was originally proposed in the work of Chay and Keyzer \cite{Chay1983} on secretory (pancreatic) cells. None of those models include slow regenerative channels, meaning that none of those models includes a source of slow negative conductance.  The fact that a minimal model of bursting does not require a slow negative conductance probably reinforces the common practice of considering instantaneous calcium activations in computational models.

If minimal models of the literature show that a slow negative conductance is not necessary to bursting, our results suggest that bursting models that lack a slow negative conductance are necessarily fragile and rigid. Robustness and tunability are not addressed in mathematical textbooks, which focus on {\it classification} rather than {\it modulation}. Starting with the seminal work of Rinzel \cite{Rinzel1985,Rinzel1987}, the mathematical theory of bursting has relied on a classification based on the possible bifurcations that can govern the transition between rest and spike. The recent work of Izhikevich \cite[page 376]{Izhikevich2007} provides up to sixteen different such mechanisms.

Our results show that robustness and tunability are properties that are distinct from a mathematical classification based on bifurcations. For instance, Figure \ref{FIG:4} illustrates that a saddle-node / saddle-homoclinic burster can be fragile or robust. And it also illustrates that a burster can be continuously modulated between different shapes without affecting the two bifurcations that determine its mathematical class. 

\subsection*{The feedback motif of robust and tunable bursting}

Our analysis of bursting modulation in terms of {\it conductances} in different time scales and different voltage ranges has a more general interpretation in terms of distinct {\it feedback} loops. When a current source has a positive conductance, it provides negative feedback to the circuit because it counteracts variations of the membrane potential. When it has a negative conductance, it provides positive feedback to the circuit because it amplifies variations of the membrane potential. With this terminology, the main message of this paper is that a minimal motif of bursting is a three feedback motif whereas a four feedback motif is required for robust modulation. The fundamental role of the slow negative conductance is interpreted as the fundamental role of a slow positive feedback.

It is common to associate excitability to a two feedback motif : a fast positive feedback for the spike upstroke and a slow negative feedback for the refractory period. The minimal motif of bursting only adds a third ultraslow negative feedback for the ultraslow adaptation between rest and spike. Instead, our results highlight that bursting is modulated by a balance of negative and positive feedbacks in the slow time scale. 

The interpretation of the results in terms of feedback loops links our results to similar findings in other areas of systems biology, see e.g. \cite{Ferrell2008}. Positive feedback loops are the essence of switches and thresholds. Our emphasis on distinct sources of fast and slow positive feedbacks has therefore the interpretation of the necessity of two rather than one thresholds for the robustness and modulation of bursting.  Each threshold accounts for two discrete states of the circuit. An excitable circuit relies on one threshold, which separates two discrete states: rest and spike. Our results suggest that a tunable bursting circuit relies on both a fast and a slow thresholds.  The two thresholds determine four discrete states: rest, tonic spiking, slow spiking, and bursting. In the absence of a slow negative conductance, the circuit has only one source of positive feedback, leading to a single threshold that makes the distinction between spiking and bursting fragile and rigid. The distinction between one and two thresholds also has a clear interpretation on the phase portraits of Figure \ref{FIG:4} . The thresholds correspond to points of ultrasensitivity where small perturbations of the initial condition leads to large differences in the resulting trajectory.  Regenerative phase portraits have distinct {\it fast} and {\it slow} thresholds. In contrast, restorative phase portraits have only one threshold, that requires a specific ratio between the fast and slow time scales.

We stress that the distinction between  four distinct feedback loops of a bursting motif do not necessarily match the physiological distinction between four distinct ionic currents. For instance, sodium channels usually provide a source of fast positive feedback through their activation and a source of slow negative feedback through their inactivation. More generally, a same current can contribute to several of the four feedback loops \cite{Franci2013,Drion2015}. But a particular modulation scenario will usually have a clear interpretation in terms of the four feedback loops. Central to this paper, the modulation from spiking to bursting will inevitably involve a balance between the slow positive feedback provided by slow regenerative channels and the slow negative feedback provided by slow restorative channels. 
The paper \cite{Drion2015} introduces the concept of {\it dynamic} input conductances to map the modulation of feedback loops to the modulation of conductance parameters in an arbitrary  conductance-based model. What is central to the message of the present paper is that the slow positive feedback is key to a feedback motif that robustly accounts for modulation from spiking to bursting.

 \subsection*{The essential role of the slow negative conductance is consistent with singularity theory}

The recent paper \cite{Franci2014} uses singularity theory \cite{Golubitsky1985} to propose a mathematical analysis of modulation in bursting models. It shows that conductance-based models that have tunable bursting capabilities have a normal form organised by a codimension three winged-cusp singularity. All the attractors that can be generated in the vicinity of this singularity can be described in a four dimensional parameter space : three unfolding parameters and the bifurcation parameter. Those abstract parameters aggregate all possible behaviors of the original model, regardless of the number of physiological parameters. They define the parameters that suffice to shape the attractors of the model.

It is a remarkable mathematical prediction from singularity theory that the four shaping parameters match the gains of the four feedback loops, or equivalently, the four distinct conductances 
discussed in the present paper. In particular, the bifurcation parameter of the model \cite{Franci2014} precisely has the interpretation of the balance between positive and negative feedback in the slow time scale. It is this parameter that governs the transition from spiking to bursting in the normal form. Singularity theory therefore identifies this one parameter as the fundamental tuning parameter of a tunable burster. This mathematical prediction was verified in six different published models  of the literature \cite{Franci2013}. A key message of the present paper is that this parameter cannot be tuned in a model that lacks currents with slow negative conductance. Modelling bursting without a slow negative conductance necessarily leads to models with  less unfolding parameters than the codimension of its organizing center, a typical {\it ground for caution} in singularity theory \cite[Section~IV.1]{Golubitsky1985}.

\end{document}